\documentclass[11pt,fleqn]{article}       

\usepackage{amsmath}
\usepackage{amsfonts}
\usepackage{graphicx}
\usepackage{dcolumn}
\usepackage{upgreek}
\usepackage{latexsym,slashed}
\usepackage{amssymb,amsfonts}
\usepackage{cancel}
\usepackage{bm}        
\usepackage{amsmath,mathrsfs}   
\usepackage{cite}

 \catcode`\@=11

\def\be{\begin{equation}}
\def\ee{\end{equation}}    
\def\beq{\begin{equation}}
\def\eeq{\end{equation}}    

\def\ba{\begin{eqnarray}}
\def\ea{\end{eqnarray}}    

\def\t{\tau}
\def\a{\alpha}
\def\l{\lambda}
\def\d{\mbox{D}}
\def\D{\mbox{Det}}
\def\R{{\hbox{{\rm I}\kern-.2em\hbox{\rm R}}}}   
\def\H{{\hbox{{\rm I}\kern-.2em\hbox{\rm H}}}}   
\def\N{{\hbox{{\rm I}\kern-.2em\hbox{\rm N}}}}   
\def\C{{\ \hbox{{\rm I}\kern-.6em\hbox{\bf C}}}} 
\def\Z{{\hbox{{\rm Z}\kern-.4em\hbox{\rm Z}}}}   
\def\ii{\infty}                                  
\def\tr{\mathop{\rm tr}\nolimits}                
\def\Tr{\mathop{\rm Tr}\nolimits}                

 \catcode`\@=11
\@addtoreset{equation}{section}

\begin{document}
\tolerance=5000

\author{ Roberto~Di~Criscienzo\thanks{rdicris@science.unitn.it} and
  Sergio~Zerbini\thanks{zerbini@science.unitn.it}\\
\\
\vspace{0.3cm}
\\
Dipartimento di Fisica, Universit\`a degli Studi di Trento \\
and Istituto Nazionale di Fisica Nucleare - Gruppo Collegato di Trento\\
Via Sommarive 14, 38100 Povo (TN), Italia
}

\title{ \bf Functional Determinants in Higher Derivative Lagrangian Theories}

\maketitle

\abstract{Motivated by the considerable success of alternative theories of gravity, we consider the toy model of a higher derivative Lagrangian theory, namely the Pais-Uhlenbeck oscillator studied in a recent paper by Hawking \& Hertog. Its Euclidean Path Integral is studied with a certain detail and a pedagogical derivation of the propagator, which makes use of a 
Theorem due to Forman,  is consequently proposed.    
}


\maketitle

\section{Introduction}

The discovery made by Riess and Perlmutter and respective collaborators  \cite{Riess:1998cb,Perlmutter:1998np} that the universe where we live is expanding with an accelerating rate, probably represents the greatest challenge of the century that Nature has provided to theoretical physicists (see, for example \cite{Padmanabhan:2006ag}). 
Among the large number of models and physical mechanisms proposed in order to explain the accelerating era of the universe, are of some interest the so-called $f(R)$-theories of gravity (see, for example\cite{review}). 
Loosely speaking, they represent alternative theories of gravity where, in place of the Einstein-Hilbert action 
\be
I_{EH} = \frac{1}{16\pi G} \int dx \sqrt{-g} R \nonumber
\ee       
the Ricci scalar is substituted by some appropriately chosen function $f(R)$ (or other higher order invariants like the 
Gauss-Bonnet invariant as in \cite{sasaki,fGB,Mota,GB,cognola06}). 
>From this perspective, the present acceleration era is a manifestation of the new, more involved, geometry of the universe. This represents a conceptual difference from models where the acceleration is driven instead by the presence of exotic 
fields (quintessence, phantom field) which dominate the matter content of spacetime. 

We should point out that the idea of introducing a correction to the Einstein-Hilbert action in the form 
of $f(R) = R + R^2$ was proposed long time ago by Starobinsky \cite{Starobinsky:1980te} in order to solve many of the problems left open by the so-called Hot Universe Scenario\footnote{This in turn had the consequence of introducing an accelerating expansion in the primordial universe, so that the Starobinsky model can be considered as the first inflationary model.}. 

Nowadays, $f(R)$-theories of gravity are understood as toy models without any intention of definiteness, that is, useful playgrounds where new physics, possibly related to observable events of our universe, can appear. Their interest grew up in 
cosmology with the appearance of the papers \cite{turner,capo}.

From a mathematical point of view, $f(R)-$theories of gravity, polynomial in the scalar curvature and with degree higher than one of course (as the Starobinsky model), are known to be asymptotically free \cite{Fradkin:1981iu} and renormalizable \cite{Stelle:1976gc,Utiyama:1962sn,Fradkin:1974df}. However, the introduction of higher derivatives of the metric seems to lead to ghosts, states with negative norm which we think are able to spoil any QFT from physical interest. A recent paper by 
Hawking \& Hertog \cite{Hawking:2001yt} has revitalized the interest in such kind of theories since, as shown by the authors, starting from a positive definite action, one can give meaning to the Euclidean path integral as a set of rules for calculating probabilities for observations. With reference to the specific model of a fourth order Lagrangian, namely that of the Pais-Uhlenbeck oscillator (PU)\cite{Pais:1950za} , it is shown that, paying the prize of losing unitarity, one can never produce or observe negative norm states. This result proves the goodness of the model and justify in our opinion the attention received recently in \cite{Andrzejewski:2009bn}. It is clear that the first step toward the comprehension of the PU quantum mechanical oscillator is represented by proper evaluation of its propagator. Both \cite{Hawking:2001yt,Andrzejewski:2009bn} have performed such calculation. Our results, modulo a normalization constant, are in agreement with the ones presented in
\cite{Andrzejewski:2009bn}, also if different techniques have been implemented. Aim of the present paper is to evaluate explicitly step by step the propagator of the PU oscillator.

\section{Path Integral Representation of PU Oscillator Propagator}

Let us consider the one-dimensional PU Lagrangian \cite{Pais:1950za},  $t \in \mathbb{R}$,
\be
L_{PU} = \frac{1}{2} \left(\frac{d q}{dt}\right)^2 - V(q) - \frac{\alpha^2}{2} \left(\frac{d^2 q}{dt^2}\right)^2\;, \qquad \mbox{with}\qquad m, \alpha >0 \;.\label{pu lagrangian}
\ee
Its Euclidean version, obtained by Wick rotating the time $t$, i.e. $t\rightarrow -i\t$, is 
\be
 - L_E = \frac{1}{2} \left(\frac{d q}{d\t}\right)^2 + V(q) + \frac{\alpha^2}{2} \left(\frac{d^2 q}{d\t^2}\right)^2\;\label{euclidean lagrangian}.
\ee
For brevity, we shall denote with an overdot the derivative with respect to $\t$. \\
Set $\hbar =1$, we can formally write the propagator as 
\be
 \int_{\mathcal{A}} \d q \, e^{i \int_0^T dt L_{PU}} \quad\stackrel{Wick}{\longrightarrow}\quad Z_T (\mathcal{A}):=\int_{\mathcal{A}} \d q\, e^{- I_E [q]}
\label{propagator}
\ee
with the Euclidean action given by
\be
I_E [q] = \int_0^T d\t \left(\frac{1}{2} \dot q^2 + V(q) + \frac{\alpha^2}{2} \ddot q^2 \right)\; \label{euclidean action}.
\ee
The Euclidean action turns out to be  positive definite as far as $V(q)$ is non negative and the propagator 
(\ref{propagator}) as explained in \cite{Hawking:2001yt} can be used, at least in the Gaussian approximation, to extract probabilities for physical  observations. Here, $\d q$ represents the formal functional measure and $\mathcal{A}$, the boundary conditions necessary to give a meaning to a formal path integral. \\
As well known, in the usual second order theory, it would be sufficient to specify $q$ on the initial and final time slice in order to make the propagator well defined. However, the present theory is of order higher than two so that extra boundary conditions are expected to be involved in the definition of (\ref{propagator}). As already proposed in \cite{kleinert}, and fully explained in \cite{Hawking:2001yt}, the right choice is provided by 
\be
\mathcal{A}: \qquad \qquad \qquad  q(0) = q_0 \;, \qquad q(T)=q_T \;, \qquad\qquad \dot q(0)= \dot q_0\;, \qquad \dot q(T) = \dot q_T \label{A1}\;,
\ee
since, any condition on $\ddot q$ would make otherwise the action infinite. 
 
Established the boundary conditions, the propagator (\ref{propagator}) can be re-written in the more evocative form
\be
Z_T (\mathcal{A}) = \langle q_T, \dot q_T; \t=T \,|\, q_0, \dot q_0; \t=0   \rangle. \label{propagator2}
\ee
Now, one may formally procedes  by splitting $q$ into a \textquotedblleft classical'' part, $q_{cl}$, and a quantum fluctuation $\hat q$, i.e. 
\be
q(\t) := q_{cl}(\t) + \hat q(\t) \label{splitting}\,.
\ee  
$q_{cl}$ solves the classical EOM obtained by $\delta I_E =0$ with boundary conditions (\ref{A1}). From (\ref{splitting}) and (\ref{A1}), it turns out that quantum fluctuations have to satisfy the following boundary conditions,
\be
\hat{\mathcal{A}}: \hspace{4.5cm} \hat q(0) =0=\hat q(T)  \qquad \&\qquad \dot {\hat q}(0)=0=\dot {\hat q}(T) \;.\label{qf bc1}
\ee
The Euclidean action (\ref{euclidean action}) becomes: 
\ba
I_E [q_{cl} + \hat q] &=& I_E [q_{cl}]  + I_E [q_{cl},\hat q] + \int_0^T d\t \,(\dot{\hat q} \dot q_{cl} + m^2 \hat q q_{cl} + \a^2\ddot{\hat q} \ddot q_{cl} ) \nonumber \\
&\stackrel{PI}{=}&   I_E [q_{cl}]  + I_E [q_{cl},\hat q] + \int_0^T d\t \,\hat q \left[\a^2 \frac{d^4 q_{cl}}{d \t^4} - \frac{d^2 q_{cl}}{d\t^2} + V'(q_{cl})\right] + \nonumber \\
& & \hspace{2.4cm} + \left(\dot q_{cl} \hat q + \a^2 \ddot q_{cl} \dot{\hat q} - \a^2 \dddot q_{cl} \hat q\right)\Big\vert_0^T\,, \label{wait}
\ea
where
\be
I_E [q_{cl},\hat q]=\int_0^T d\t \left(\frac{1}{2} \dot{\hat q}^2  + \frac{\alpha^2}{2} \ddot{\hat  q}^2 +
W(q_{cl},\hat q) \right)\;. 
\label{euclidean action testa}
\ee
with
\be
W(q_{cl}, \hat q)=V(q_{cl}+\hat q)-\hat q V'(q_{cl})- V(q_{cl})\,.
\label{w}
\ee
Notice that, by extremizing the action, we get the EOM for the classical solution, namely.
\be
\mbox{EOM:} \hspace{4.5cm} \a^2 \frac{d^4 q_{cl}}{d \t^4} - \frac{d^2 q_{cl}}{d\t^2} + V'(q_{cl}) =0 \label{EOM}\,.
\ee
In (\ref{wait}) the integral vanishes on-shell, while the boundary term vanishes upon using (\ref{qf bc1}). To go on, we make the usual Gaussian approximation or equivalently, we may restict to quadratic potentials of the type $V(q)=\frac{m^2}{2}q^2$. Thus,  $W(q_{cl}, \hat q) = W (\hat q)= \frac{m^2}{2} \hat q^2$, with $m^2$ constant. The Euclidean action after the splitting (\ref{splitting}) and in reason of this consideration neatly separates into
\ba
I_E[q] &=& I_E[q_{cl}] +  I_E [\hat q] \nonumber \\
&\stackrel{(\ref{qf bc1})}{=}&I_E[q_{cl}] + \frac{1}{2}\int_0^T d\t \;\hat q(\t) \left[\a^2 \frac{d^4}{d \t^4} - \frac{d^2}{d\t^2} +m^2\right] \hat q(\t) \, \label{step}
\ea
and the classical action has been explicitly evaluated in \cite{Hawking:2001yt}.\\
The propagator assumes the form of a Gaussian integral in the quantum fluctuation variables:
\ba
& &\langle q_T, \dot q_T; \t=T \,|\, q_0, \dot q_0; \t=0   \rangle \stackrel{(\ref{step})}{=} \exp -I_E[q_{cl}] \times \nonumber \\
& & \hspace{3.5cm}\times \int_{\mathcal{A}} \d \hat q \,\exp -\frac{1}{2}\int_0^T d\t \;\hat q(\t) \left[\a^2 \frac{d^4}{d \t^4} - \frac{d^2}{d\t^2} +m^2\right] \hat q(\t) \label{gaussian}
\ea
\\
\\
As a consequence, one has to give a meaning to the formal Gaussian path integral. To this aim, let us denote 
\be
K_\mathcal{X} := \a^2 \frac{d^4}{d \t^4} - \frac{d^2}{d\t^2} + m^2 \;,\label{K}
\ee
as the fourth-order differential operator in $L_2(0,T)$, defined in the dense domain  
$D(K) := \{f, K_\mathcal{X} f \in L_2 (0,T) \;|\; \mathcal{X} (f) = 0 \}$  with suitable boundary conditions $\mathcal{X}$.\\
Let us try to determine $\mathcal{X}$ such that $K$ is a self-adjoint operator in  $L_2(0,T)$. 
To this purpose, suppose $f^{(n)}$ ($n=0,\dots,4$) be in $D(K)$:
\ba
(g,K f) &:=& \int_0^T d\t \bar g(\t) \left[\a^2 \frac{d^4}{d \t^4} - \frac{d^2}{d\t^2} +m^2\right] f(\tau) \nonumber \\
&=& \int_0^T d\t  \left[\a^2 \frac{d^4}{d \t^4} - \frac{d^2}{d\t^2} +m^2\right] \bar g(\t) f(\t) + \nonumber\\
& & + \dot{\bar g} f - \bar g \dot f + \a^2 (\bar g \dddot f - \dot{\bar g} \ddot f + \ddot{\bar g} \dot f - \dddot{\bar g} f) \Big\vert_0^T \nonumber \\
&=& (Kg,f) + B(0,T) .
\ea  
$K$ is certainly symmetric  if and only if
\be
B(0,T) = 0 .
\ee
This condition is realized, among the others, by the functions $f^{(n)}(\t)$, $g^{(n)}(\t)$ ($n=0, \dots,4$) absolute continuous in $[0,T]$ s.t.  
\be
 f(0)=0=f(T) \quad \& \quad \dot f(0)=0=\dot f(T) \;;\quad \quad  g(0)=0=g(T) \quad \& \quad \dot g(0)=0=\dot g(T) \label{BC1} 
\ee
or 
\be
f(0)=0=f(T) \quad \& \quad \ddot f(0)=0=\ddot f(T) \;; \quad\quad g(0)=0=g(T) \quad \& \quad \ddot g(0)=0=\ddot g(T)  \label{BC2} \;.
\ee
Thus, we may invoke the general Von-Neuman-Krein method to find all the self-adjoint extensions of a symmetric operator. 
In our case, one has to find the $L_2(0,T)$ solutions to equation
\be
K^\dag u_{\pm}(\tau)=\pm i u_{\pm}(\tau)\,.
\ee
Since  $(0,T)$ is compact, it turns out that the defect indices are $(n_+,n_-)=(4,4)$  meaning that all  
self-adjoint extensions are parametrized by $4 \times 4$ unitary matrices. For our pourposes, we simply observe that physical boundary conditions (\ref{BC1}), corresponding to  (\ref{A1}) - (\ref{qf bc1}), 
can be represented by an unitary matrix, and the operator $K$ is self-adjoint. 
This is also confirmed by the fact that the associated spectral problem is well defined,
eventhough the equation which implicitly defines the eingenvalues is highly trascendental. In fact,  for example, even the 
spectrum of the simplest fouth order operator $K^{(0)}:= \a^2\frac{d^4}{d\t^4}$ with boundary conditions  (\ref{BC1}) , 
is analytically unaccessible due to the impossibility of solving the trascendent equation\footnote{Said $\l_n$ the $n$-th eigenvalue of $K^{(0)}$, then $z_n(\l_n):=\sqrt{\a} \l_n^{1/4}$ .} $\cos(T z_n) = \mbox{sech}(T z_n )$. 
\vspace{0.4cm}\\
On the other hand, it is also easy to show that the boundary conditions  (\ref{BC2}) defines also a self-adjoint 
extension. But in this case, the spectral problem
is much more easier to handle. In fact, let us denote by 
$\bar K$ 
\be
\bar K = L_0 + m^2 + \a^2 L_0^2 \;, \qquad \mbox{with (\ref{BC2}) as BC}\,,
\ee
where $L_0= - \frac{d^2}{d \tau^2}$. The eigenfunctions are $\sin\left(\frac{\pi\t}{T}\right)$ and the the spectrum is 
\be
\sigma(\bar K):= \{\bar\l_n := \left(\frac{\pi n}{T}\right)^2 + m^2 + \a^2 \left(\frac{\pi n}{T}\right)^4, \, n\in N\}\;.\label{unphys spectrum}
\ee
The problem is that boundary conditions  (\ref{BC2}) are ``unphysical'', in the sense that they do not enter in the
expression of our propagator. Nevertherless, as we will see, they will play an important role.

We conclude this Section with the final form of PU propagator, obtained
performing the Gaussian integral (\ref{gaussian}):
\be
\langle q_T, \dot q_T; \t=T \,|\, q_0, \dot q_0; \t=0   \rangle = \sqrt{\frac{2\pi}{\D\,K_{\mathcal{A}}}} \,\exp \left(-I_E[q_{cl}]\right) \;. \label{evaluation}
\ee   

\section{Regularization of Functional Determinants}

The goal is to give a rigourous meaning to  the formal functional determinant which appears in (\ref{evaluation}).  
This is also called prefactor.
In \cite{Hawking:2001yt}, the authors  have performed the calculation of the prefactor under the (implicit) 
assumption that the Van-Vleck-Pauli method is valid even for the higher-order dynamical system. 
This is not obvious, since the Pauli theorem is based on Hamilton-Jacobi theory for ondinary second order Lagrangian. In our case, the Hamiltonian formalism is quite different from the ordinary one, known as Ostrogadsky formulation.
A second attempt has been recently proposed by \cite{Andrzejewski:2009bn} but we reserve to comment it later.
\vspace{0.6cm}\\
In our approach, we shall regularize the functional determinant by zeta-function regularization 
\cite{z1,z2,dowk76-13-3224,eli94,byts96}. A regularization is necessary since 
 functional determinants are formally divergent. We recall that a simple regularization for the functional determinant 
associated with  an elliptic operator $L$ may be chosen as
\beq
\ln\, \D\, L (\varepsilon)=-\int_0^\infty dt\  \frac{t^{\varepsilon-1}}{\Gamma(1+\varepsilon)} 
\Tr e^{-tL/\mu^2} =-\frac{1}{\varepsilon}\zeta\left(\varepsilon\,\Big\vert \frac{L}{\mu^2}\right)\,,
\label{bb}
\eeq
where the zeta-function is defined by means of the Mellin-like transform 
\beq
\zeta(s|L)=\frac{1}{\Gamma(s)}\int_0^\ii dt\ t^{s-1} \Tr e^{-tL}\qquad \& \qquad \zeta\left(s\, \Big\vert \frac{L}{\mu^2}\right) =\mu^{2s}\,\zeta(s|L)\:.
\label{mt}\eeq
For a $Q$ order differential operator in D-dimensions, 
the integral is convergent in the region $\mbox{Re}\, s> \frac{D}{Q}$ where the function $\zeta(s|L)$ is analytic. 
It is possible to show that $\zeta(s|L)$ can be analytically continued in 
the whole complex plane and it is regular at $s=0$. Thus, by expanding (\ref{bb}) in Taylor's series, we obtain
\beq
\ln\, \D\,L(\varepsilon)=-\frac{1}{\varepsilon} \zeta\left(0\, \Big\vert\frac{L}{\mu^2}\right) -\zeta'\left(0\, \Big\vert\frac{L}{\mu^2}\right) + \mathcal{O}(\varepsilon)
\label{dow33}
\eeq
and the regularised functional determinant associated with $L$ can be  defined
by taking the finite part in the limit $\varepsilon \to 0$,
that is  
\beq 
\ln\,\D\, L = -\zeta'\left(0\, \Big\vert\frac{L}{\mu^2}\right) = -\zeta'(0|L) -\ln\mu^2\:\zeta(0|L)\,,
\eeq
leading to the usual zeta-function regularisation prescription \cite{z1,z2}. The divergences are governed by the computable coefficient
$\zeta(0|L)$, which does not depend on the 
arbitrary scale parameter $\mu$.\\
\\
Within this direct zeta-function regularization, in order to calculate $\D K_{\mathcal{A}}$, we need to know explicitly
the spectrum.  As alreday stressed, often one does not  know explicitly the spectrum. Thus, we are forced to  
 make use of a powerful theorem proved by Forman in \cite{Forman:1987}, which is a generalization of Gelfand-Yaglom 
\cite{GY} and Levit-Smilansky theorems Levit-Smilansky theorems \cite{LS}. Adapted to the case at end, we may state it in the following way\footnote{What we report here is a simplified version of \underline{Proposition 3.9} in \cite{Forman:1987}.}:
\vspace{0.6cm}
\begin{flushleft} 
\textbf{Theorem} (Forman, 1987):   Let $K$, $\bar K$ be operators defined in $[0,T]$ of the form
\ba
K &=& P_0(\t) \frac{d^n }{d\t^n} + \mathcal{O}\left(\frac{d^{n-2} }{d\t^{n-2}}\right) \nonumber\\
\bar K &=& P_0(\t) \frac{d^n }{d\t^n} + \mathcal{O} \left(\frac{d^{n-2} }{d\t^{n-2}}\right)  . 
\ea
Let $\mathcal{A}$ be any boundary condition represented by matrices $(M,N)$, s.t.
\be
M \left(\begin{array}{c}
h(0)  \\
\cdots \\
h^{(n-1)}(0)
\end{array}\right) + N \left(\begin{array}{c}
h(T)  \\
\cdots \\
h^{(n-1)}(T)
\end{array}\right) = 0 \;.
\ee
Then,
\be
\frac{\D K_\mathcal{A}}{\D \bar K_\mathcal{\bar A}} = \frac{\det (M + N Y_{K}(T))}{\det (\bar M + \bar N Y_{\bar K}(T))} \;.\label{forman}
\ee 
where  $(\bar M,\bar N)$ are matrices defining other boundary conditions smoothly related to $(M,N)$.
Above, for any $h$ such that $K h =0$, $Y_K(\t)$ acts as 
\be
\left(\begin{array}{c}
h(\t)  \\
\cdots \\
h^{(n-1)}(\t)
\end{array}\right) = Y_K(\t) \left(\begin{array}{c}
h(0)  \\
\cdots \\
h^{(n-1)}(0)
\end{array}\right) \;. \label{Y}
\ee
\end{flushleft}
\vspace{0.6cm}
A solution of this equation is given by 
\be
Y_K(\t) = \left(\begin{array}{cccc}
u_1(\t) & u_2(\t) & u_3(\t) & u_4(\t)\\
\dot u_1(\t) & \dot u_2(\t) & \dot u_3(\t) & \dot u_4(\t) \\
\ddot u_1(\t) & \ddot u_2(\t) & \ddot u_3(\t) & \ddot u_4(\t) \\
\dddot u_1(\t) & \dddot u_2(\t) & \dddot u_3(\t) & \dddot u_4(\t) 
\end{array}\right) \;, 
\ee 
where $K u_j=0$ for $j=0,\dots 4$, satisfying the initial conditions
\ba
u_1(0)&=&1 \;, \qquad u_j(0)=0 \;,\quad j\neq 1 \;, \nonumber \\
\dot u_2(0)&=&1 \;,  \qquad \dot u_j(0)=0 \;, \quad j\neq 2 \;,\nonumber \\
\ddot u_3(0)&=&1 \;,  \qquad \ddot u_j(0)=0 \;, \quad j\neq 3\;,\nonumber \\
\dddot u_4(0)&=&1 \;, \qquad \dddot u_j(0)=0 \;, \quad j\neq 4 \label{u BC} .
\ea
\vspace{0.4cm}\\
The role of $K$ which appears in the Theorem is played by the operator (\ref{K}). The boundary conditions (\ref{qf bc1}) can be put in correspondence with matrices\footnote{The choice is highly non unique of course. The final result, however, seems not to depend on this large freedom. Cfr. \cite{Forman:1987}.} 
\be
M= \left(\begin{array}{cc}
\mathbb{I}_{(2)} & 0 \\
0 & 0
\end{array}\right) \;, \qquad N= \left(\begin{array}{cc}
0 & 0 \\
\mathbb{I}_{(2)} & 0
\end{array}\right) .\label{MN}
\ee  
Finally, in our case, the matrix $Y_{K}(T)$ can be easily computed. The general solution of $K u=0$ reads
\be
u_j(\t) = A_j \sinh(\l_1 \t) + B_j \cosh(\l_1\t) + C_j \sinh(\l_2\t) + D_j \sinh(\l_2\t) \,,\label{fundamental soln}
\ee
with
\be
\l_{1,2} = \sqrt{\frac{1\mp \sqrt{1-4\a^2 m^2}}{2\a^2}} \label{lambda} \;.
\ee
In the oscillatory regime, that is $2\a m< 1$, the roots $\l_{1,2}$ are real as can be checked expanding the double radicals above.\\
Imposing the initial conditions (\ref{u BC}), one has

\ba
u_1(\t) &=& \frac{\l_2^2}{\l_2^2-\l_1^2} \cosh(\l_1\t) - \frac{\l_1^2}{\l_2^2-\l_1^2} \cosh(\l_2\t) \;,\nonumber \\
u_2(\t) &=& \frac{\l_2^2}{\l_1(\l_2^2-\l_1^2)} \sinh(\l_1\t) - \frac{\l_1^2}{\l_2(\l_2^2-\l_1^2)} \sinh(\l_2\t) \;, \nonumber\\
u_3(\t) &=& - \frac{1}{\l_2^2-\l_1^2} \cosh(\l_1\t) + \frac{1}{\l_2^2-\l_1^2} \cosh(\l_2\t) \nonumber \\
u_4(\t) &=& - \frac{1}{\l_1(\l_2^2-\l_1^2)} \sinh(\l_1\t) + \frac{1}{\l_2(\l_2^2-\l_1^2)} \sinh(\l_2\t) . \label{u}
\ea
Therefore, the right hand side numerator of (\ref{forman}) is 
\ba
& & \det (M + N Y_K(T)) \stackrel{(\ref{MN})}{=} u_3(T) \dot u_4(T) - u_4(T) \dot u_3(T) \hspace{6cm} \nonumber \\
& & \hspace{1cm} \stackrel{(\ref{u})}{=} \frac{1}{(\l_2^2-\l_1^2)^2} \left(2-2\cosh(\l_1 T) \cosh(\l_2 T) + \frac{\l_1^2+\l_2^2}{\l_1\l_2} \sinh(\l_1 T) \sinh(\l_2 T)\right) \nonumber \\
& & \hspace{1cm}\stackrel{(\ref{lambda})}{=} \frac{\a^3}{m} \left[\frac{1}{1+2\a m} \sinh^2\left(\frac{\sqrt{1+2\a m} T}{2\a}\right)  - \frac{1}{1-2\a m} \sinh^2\left(\frac{\sqrt{1-2\a m} T}{2\a}\right) \right] .
\label{num}
\ea
In order to compute completely the propagator, we need to find the fourth-order differential operator which will play the role of $\bar K$ in Forman's theorem. The requirement on the candidate is that we must be able to compute its funtional determinant in an independent way. Loosely speaking, there are two ways to compute the functional determinant of an operator, that is: (i) taking the product of its eigenvalues (if you know the spectrum); (ii) using some \textquotedblleft smart'' mathematical theorem. We have already decided to follow the latter to compute $\D \,K$, something which forces us to take the former way for $\D \,\bar K$. \\
\\
As we have seen, Forman's theorem lets us to play with the boundary conditions both of $K$ and $\bar K$. This suggests us to choose $\bar K$ formally equal to $ K$, with unphysical boundary conditions (\ref{BC2}). We recall that in this case we known the spectrum, given in (\ref{unphys spectrum}).\\
A quick way to compute the functional determinant $\D\, \bar K$ may be the following. 
Zeta-funtion regularized determinants suffer, in general of the so-called 
multiplicative anomaly \cite{eliz98-194-613,byts98-38-1075,eliz98u-413}, namely
\be
\D (\,A \,B)=\D \,A \;\D \,B\ e^{a(A,B)}\,. 
\ee
The quantity$a(A,B)$ is called multiplicative anomaly. In our case, by noticing that
\be
\bar K = \left(\a L_0 + \frac{1 - \sqrt{1-4\a^2 m^2}}{2\a}\right)\left(\a L_0 + \frac{1 + \sqrt{1-4\a^2 m^2}}{2\a}\right) \equiv \bar K_- \bar K_+\;,
\ee
and observing that,  since $\bar K$ is an ordinary fourth order differential operator, the multiplicative anomaly 
is vanishing (cf. \cite{eliz98u-413}), we have
\be
\D\,\bar K = (\D\,\bar K_-)(\D\,\bar K_+).
\ee
In this way, $\bar K_\pm$ are just second order differential operators obtained by shifting the simple $L_0$ 
with a constant term. The calculation of  $(\D\,\bar K_{\pm})$ is standard and we are not going to repeat it here. The result is 
\be
\D\,\bar K = \frac{\a}{m T^2}  \left[\sinh^2\left(\frac{\sqrt{1+2\a m} T}{2\a}\right) - \sinh^2\left(\frac{\sqrt{1-2\a m} T}{2\a}\right) \right] \label{Kbarra} \;.
\ee
We may confirm this result (and in turn, the absence of multiplicative anomaly) by evaluating directly the 
determinant of $\bar K$. This target can be accomplished by implementing a standard trick which consists in the observation that the derivative with respect to $m^2$  of $\ln (\D\,\bar K)$ is a well defined quantity, namely
\ba
\frac{d}{d(m^2)} \ln (\D\,\bar K) &=& \frac{d}{d(m^2)} \tr (\ln \bar K)\nonumber\\
&=& \frac{d}{d(m^2)} \sum_{n=1}^\infty \ln(\bar\l_n) \nonumber \\
&=& \sum_{n=1}^\infty \frac{1}{\bar\l_n}  < \infty \label{s}
\ea 
Thus, the problem has been reduced to compute the convergent series
\ba
 \mathcal S &\equiv&  \sum_{n=1}^\infty \frac{1}{\bar\l_n}  \nonumber \\
&=&\sum_{n=1}^\infty \frac{1}{\left(\frac{\pi n}{T}\right)^2 + m^2 + \a^2 \left(\frac{\pi n}{T}\right)^4} \nonumber \\
&=& \frac{T^2}{\sqrt{1-4\a^2 m^2}} \left(\sum_{n=1}^\infty  \frac{1}{(\pi n)^2 + z_-^2}  -\sum_{n=1}^\infty\frac{1}{(\pi n)^2 + z_+^2}\right) \;,
\ea
where for simplicity we have introduced the quantities
\be
z_\pm := T \sqrt{\frac{1\pm \sqrt{1-4\a^2 m^2}}{2\a^2}}  = \frac{T}{2\a} \left(\sqrt{1+2\a m} \pm \sqrt{1-2\a m}\right).
\ee
In terms of these new quantities,
\be
d(m^2) = \mp \frac{\sqrt{1-4\a^2 m^2}}{T^2} (2z_\pm) dz_\pm.
\ee
Recalling the Mittag-Leffler expansion of the $\coth$ function, i.e.
\be
\coth z = \frac{1}{z} + 2z  \sum_{n=1}^\infty \frac{1}{(\pi n)^2 + z^2},
\ee
it turns out that
\ba
\ln (\D\,\bar K) &=& \int dz_- \left(\coth(z_-)- \frac{1}{z_-}\right) +  \int dz_+ \left(\coth(z_+)- \frac{1}{z_+}\right) \nonumber \\
&=& \ln \left(\frac{\sinh(z_+)\sinh(z_-)}{z_+z_-}\right)\,,
\ea
so that the final result coincides indeed with (\ref{Kbarra}).
\vspace{0.6cm}\\
Now that we know $\D\,\bar K$ from independent considerations, we can close the circle just computing the denominator of (\ref{forman}). Since $K$ and $\bar K$ are formally the same, the $Y$ matrix does not change. What changes are the boundary conditions which now are represented, for example, by
\be
\bar M= \left(\begin{array}{cccc}
1 & 0 & 0 & 0 \\
0 & 0 & -1 & 0\\
0 & 0 & 0 & 0 \\
0 & 0 & 0 & 0
\end{array}\right) \;, \qquad \bar N= \left(\begin{array}{cccc}
0 & 0 & 0 & 0 \\
0 & 0 & 0 & 0 \\
1 & 0 & 0 & 0 \\
0 & 0 & 1 & 0
\end{array}\right) .\label{MNbar}
\ee  
The right hand side denominator of (\ref{forman}) becomes
\ba
& & \det (\bar M + \bar N Y_{\bar K}(T)) = u_2(T)\ddot u_4(T) - u_4(T) \ddot u_2(T) \hspace{6.5cm}\nonumber\\
& &\hspace{1.4cm} \stackrel{(\ref{u})}{=} \frac{1}{\l_1\l_2} \sinh(\l_1 T)\sinh(\l_2 T) \nonumber \\
& & \hspace{1.4cm}\stackrel{(\ref{lambda})}{=} \frac{\a}{m} \cdot \left[\sinh^2\left(\frac{\sqrt{1+2\a m} T}{2\a}\right) - \sinh^2\left(\frac{\sqrt{1-2\a m} T}{2\a}\right) \right] \label{den}.
\ea
Inserting (\ref{den}), (\ref{num}) and (\ref{Kbarra}) into (\ref{forman}), the functional determinant of the $K$ operator with boundary conditions (\ref{A1}) - (\ref{qf bc1}) is 
\ba
\D\,K_{\mathcal{A}} &=& \frac{\a^3}{m T^2} \left[\frac{1}{1+2\a m} \sinh^2\left(\frac{\sqrt{1+2\a m} T}{2\a}\right) + \right. \nonumber \\
& & \left. \hspace{4.5cm} - \frac{1}{1-2\a m} \sinh^2\left(\frac{\sqrt{1-2\a m} T}{2\a}\right) \right]. \label{denK}
\ea
This is the main result of the paper. A dimensional analysis tells us that, in natural units, $[\a] = [T] =(\mbox{mass})^{-1}$, so that $[\D \,K] = (\mbox{mass})^{-2}$. Up to a minor adjustment of the numeric factor, (\ref{denK}) is the prefactor presented  in
 \cite{Andrzejewski:2009bn}, equation (9).\\ 
It is straightforaward to check that, in the small $T$ limit,
\be
\D\,K_{\mathcal{A}} \approx \frac{T^2}{12} + \frac{T^4}{180\a^2} + \mathcal{O}(T^6) \;,\qquad T\ll1\;,
\ee
but for $\t \in [0, \infty)$,
\ba
\D\,K_{\mathcal{A}} &=& \frac{\a}{m} \left[\frac{1}{1+2\a m} \,\frac{\exp\left(\sqrt{1+2\a m} \frac{T}{\a}\right)}{(2T/\a)^2} \,+\right. \nonumber \\
& & \left. \hspace{3.1cm} - \frac{1}{1-2\a m} \,\frac{\exp\left(\sqrt{1-2\a m} \frac{T}{\a}\right)}{(2T/\a)^2} \right] \,, \qquad T\gg 1\;,
\ea
something which shows that the ground state probability amplitude, being proportional to $\D\,K_{\mathcal A} ^{-1}$, is indeed exponentially suppressed and makes the Euclidean theory well defined.\\
Finally, notice that the vanishing "mass'' term case is 
\be
\D\,K_{\mathcal{A}}\; \stackrel{m\rightarrow 0}{\longrightarrow} \;\frac{\a^4}{T^2} \left[2\left(1-\cosh\left(\frac{T}{\a}\right)\right) + \frac{T}{\a} \sinh\left(\frac{T}{\a}\right)\right] =: \D\, K^{(1)}_{\mathcal{A}}\;,\label{denK1}
\ee
where we have defined the operator $K^{(1)}_{\mathcal{A}}:= \a^2 \frac{d^4}{d\t^4} - \frac{d^2}{d\t^2}$ with boundary conditions (\ref{A1}) - (\ref{qf bc1}). Given (\ref{denK1}), one can evaluate, for example, the determinant of the most elementary fourth-order differential operator of physical interest, namely $K^{(0)}_{\mathcal{A}}:= \a^2 \frac{d^4}{d\t^4}$ in $\t \in [0,T]$ (operator whose spectrum, as said above, is unaccessible to us), which is
\be
\D\,K^{(0)}_{\mathcal{A}} = \frac{T^2}{12} \label{detK0}\,.
\ee
We leave the details of the computation to the interested reader.

\section{Conclusions}

To summarize, we have devoted the paper to the computation of the quantum mechanical Euclidean propagator of the one-dimensional PU oscillator. As we have seen, the PU Lagrangian gives rise to a fourth order differential operator whose functional determiant, $\D\,K_{\mathcal{A}}$, has represented the main difficulty along the way. As showed by Hawking \& Hertog, the correct boundary conditions $\mathcal{A}$ we have to impose in order to give meaning to the Euclidean propagator, are provided by equation (\ref{A1}). By making use of Forman's theorem, we have been able to evaluate $\D \,K_{\mathcal{A}}$, so that the main result of the paper can be stated in the following way,
\be
\langle q_T, \dot q_T; \t=T \,|\, q_0, \dot q_0; \t=0  \rangle = \sqrt{\frac{2\pi}{\D\,K_{\mathcal{A}}}} \,\exp \left(-I_E[q_{cl}]\right) \;,
\ee   
where
\ba
\D\,K_{\mathcal{A}} &=& \frac{\a^3}{m T^2} \left[\frac{1}{1+2\a m} \sinh^2\left(\frac{\sqrt{1+2\a m} T}{2\a}\right) + \right. \nonumber \\
& & \left. \hspace{4.5cm} - \frac{1}{1-2\a m} \sinh^2\left(\frac{\sqrt{1-2\a m} T}{2\a}\right) \right]
\ea
and the classical Euclidean action can be found in \cite{Hawking:2001yt}, equations (A5)-(A6).
\vspace{0.4cm}\\
As a byproduct, we have shown how Forman's theorem can be usefully implemented to give other functional determinants of potential interest. To this regard, Forman's theorem proves to be an instrument of extraordinary power.
The generalization to higher order quadratic Lagrangian $L(q, q^{(1)},\cdots, q^{(r)})$ is immediate, eventhough the computations become of course much more involved. \\
We conclude with the observation that the method described in this paper may also be useful in the so-called Ho\v{r}ava-Lifshitz
non-relativistic renormalizzable theory of gravity \cite{hora}, where higher spatial derivatives appear indeed in the Lagrangian of the theory and in inflationary cosmology \cite{tim}.

\end{document}